\documentclass[10pt, journal]{IEEEtran}
\usepackage{graphicx}
\usepackage{xcolor}
\IEEEoverridecommandlockouts
\usepackage{cite}
\usepackage{amsmath,amssymb,amsfonts}
\usepackage{algorithmic}
\usepackage{graphicx}
\usepackage{multirow}
\usepackage{tikz}
\usepackage{array}
\newcolumntype{L}[1]{>{\raggedright\let\newline\\\arraybackslash\hspace{0pt}}m{#1}}
\newcolumntype{C}[1]{>{\centering\let\newline\\\arraybackslash\hspace{0pt}}m{#1}}
\newcolumntype{R}[1]{>{\raggedleft\let\newline\\\arraybackslash\hspace{0pt}}m{#1}}

\graphicspath{ {figure_new/} } 
\usepackage{textcomp}

\usepackage{xcolor}
\newcommand{\RNum}[1]{\uppercase\expandafter{\romannumeral #1\relax}}
\def\BibTeX{{\rm B\kern-.05em{\sc i\kern-.025em b}\kern-.08em
    T\kern-.1667em\lower.7ex\hbox{E}\kern-.125emX}}
\begin{document}

\title{PODTherm-GP: A Physics-based Data-Driven Approach for Effective Architecture-Level Thermal Simulation of Multi-Core CPUs  }

\author{Lin~Jiang,
        Anthony~Dowling,
        Ming-C.~Cheng,
        Yu~Liu%

\thanks{Lin Jiang, Anthony Dowling, Ming-C. Cheng and Yu Liu are with the Department
of Electrical and Computer Engineering, Clarkson University, Potsdam,
NY, USA 13699-5720.\protect\\}

\thanks{E-mail: \{jiangl2, dowlinah, mcheng, yuliu\}@clarkson.edu.\protect\\}
\thanks{This work is supported by National Science Foundation under Grant Nos. ECCS-2003307 and OAC-2118079.\protect}
}

\maketitle
\begin{abstract}
 A thermal simulation methodology derived from the proper orthogonal decomposition (POD) and the Galerkin projection (GP), hereafter referred to as PODTherm-GP, is evaluated in terms of its efficiency and accuracy in a multi-core CPU. The GP projects the heat transfer equation onto a mathematical space whose basis functions are generated from thermal data enabled by the POD learning algorithm. The thermal solution data are collected from FEniCS using the finite element method (FEM) accounting for appropriate parametric variations. The GP incorporates physical principles of heat transfer in the methodology to reach high accuracy and efficiency. The dynamic power map for the CPU in FEM thermal simulation is generated from gem5 and McPACT, together with the SPLASH-2 benchmarks as the simulation workload. It is shown that PODTherm-GP offers an accurate thermal prediction of the CPU with a resolution as fine as the FEM. It is also demonstrated that PODTherm-GP is capable of predicting the dynamic thermal profile of the chip with a good accuracy beyond the training conditions. Additionally, the approach offers a reduction in degrees of freedom by more than 5 orders of magnitude and a speedup of 4 orders, compared to the FEM.
\end{abstract}


\section{Introduction}
Thermal issues have been the bottleneck of performance improvements for high-performance microprocessors due to drastic minimization of the semiconductor technology nodes and introduction of multi-core architectures in the last several decades, which have resulted in significant enhancement of the power density in the processors~\cite{9218977}. High temperature gradients and hot spots not only impair performance of processors but also degrade their reliability~\cite{heinig2014thermal,zhou2018thermal}. Thermal management and thermal-aware design exploration of the high-performance processors~\cite{9401414,revathi2021multistep}  have been the effective approaches to minimize these thermal issues to improve the performance and reliability of the processors. For instance, as found in ~\cite{kim2020adaptive}, the average performance is further improved by 8.9\% for the heterogeneous multi-core processor through an adaptive thermal management framework, compared to ARM’s DVFS (Dynamic Voltage Frequency Scaling)-based intelligent power allocation. These effective thermal managements however require an effective thermal simulation tool. For real time applications, such as run-time thermal aware task scheduling, a very efficient thermal simulation with a high accuracy is desirable.

For thermal simulations of semiconductor chips, many approaches have been developed for different applications. Some of the approaches focus on the accuracy of the thermal simulation to capture hot spots, for instance the direct numerical simulations (DNSs) based on the finite difference, finite volume or finite element method (FDM, FVM or FEM, respectively). These DNS methods are however computationally intensive and in general prohibitive for thermal simulations at the architecture level. Several other approaches therefore have been proposed in situations where the efficiency plays an important role, including the thermal circuit model, the Green’s function method, machine learning based methods, etc. All these approaches achieve higher efficiency than DNSs by sacrificing accuracy and/or resolution with approximations that impose severe limitations. For instance, efficient thermal circuits are realized at the cost of very low resolution and inaccurate solutions. When using the thermal circuit model or the machine learning based method, if one intends to maintain resolution as fine as DNSs, intensive computational efforts become similar to DNSs. Additionally, it is difficult to apply the Green’s function method to 3D dynamic thermal simulations. In recent years, the data-driven approach based on proper orthogonal decomposition (POD), together with the guidance of physical principles, has become increasingly attractive for thermal simulation of semiconductor chips due to its ability to achieve high accuracy, resolution and efficiency simultaneously ~\cite{jiang2022ISCAS,jia2022methodology,jia2014thermal}. These thermal simulation approaches are further discussed in Sec.~\RNum{2}.

In this work, an architecture-level thermal simulator has been developed based on the POD-Galerkin (PODTherm-GP) methodology for 3D dynamic thermal simulation of a multi-core CPU. {The early concept of this methodology was briefly illustrated in~\cite{jiang2022ISCAS}.} This POD-Galerkin modeling technique offers an accurate and efficient prediction of the thermal profile in the multi-core processor without a priori assumptions. 
The POD projects the thermal problem from a physical domain of the multi-core CPU onto a functional space, whose basis functions (or POD modes) are trained by thermal data generated by DNSs of the multi-core CPU. In our study, the thermal data is collected from FEniCS, an open-source computing platform for solving partial differential equations (PDEs) using the FEM~\cite{o2}. 
To provide realistic heat sources in FEniCS-FEM thermal simulation, the power trace in each unit is generated by gem5~\cite{GEM5_2011} and McPAT~\cite{MCPAT_2009} with selected benchmarks. The validation of PODTherm-GP was conducted using a different power trace in each unit from that used for thermal data collection. 

The key concepts and contributions of the POD-Galerkin methodology are summarized as follows:
\begin{enumerate}
\item The POD process trains an optimal set of modes to acquire essential information embedded in the collected data. This ensures an accurate prediction using the smallest number of modes (i.e., DoF) to reach high accuracy if the data quality is sufficient.
\item The Galerkin projection enforces the physical principles, guided by the heat transfer equation, in the prediction and offers the extrapolation capability with good accuracy. 
\item PODTherm-GP thus offers accurate 3D dynamic thermal simulations with high efficiency and accuracy, and the thermal resolution is as fine as the training data collected via DNSs.
\item PODTherm-GP reduces numerical degrees of freedom (DoF) by more than 5 orders of magnitude with a high accuracy, resulting in a 4-order speedup, compared with FEniCS-FEM.
\item The quality of the POD modes significantly depends on the accuracy of the thermal gradients in collected training data.
\end{enumerate}

The rest of this work is organized as follows.  Firstly, the related work of the thermal simulation for semiconductor chips and processors is presented in Sec.~\RNum{2}. The PODTherm-GP methodology is introduced in Sec.~\RNum{3} from the mathematical perspective. Sec.~\RNum{4} illustrates the evaluation of the methodology. Next, PODTherm-GP is employed to perform dynamic thermal simulations of the selected multi-core CPU in Sec.~\RNum{5}, and the results are discussed in Sec.~\RNum{6} in terms of the accuracy and efficiency. Finally, the findings of this work are concluded in Sec.~\RNum{7}.

\section{related work on thermal simulation of semiconductor chips}
Thermal simulation is an important component for thermal ~\cite{kim2020adaptive,sridhar20103d} and power ~\cite{bogdan2013dynamic,cocskun2011thermal} managements which are two major strategies to minimize the coupling effects between chip temperature and power consumption and to prevent semiconductor chips from overheating. Hot spots in semiconductor chips can be predicted through accurate thermal simulation, and they can then be suppressed via the power management techniques, such as DVFS~\cite{kim2020adaptive}. The reduced temperature in turn improves the power efficiency of semiconductor chip. In addition, the chip thermal profile predicted by thermal simulation is needed in thermal management to improve chip performance and reduce power consumption through exploring the cooling techniques~\cite{cocskun2011thermal,sridhar20103d} or thermal-aware task scheduling~\cite{zhou2018thermal}.  The existing thermal modeling approaches for semiconductor chips are briefly discussed as follows.  
\subsection{Direct  Numerical Simulations (DNSs)}
 {To obtain detailed thermal profiles in semiconductor chips, usually the FEM, FVM or FDM is applied. For example, 3D thermal ADI based on the FDM was developed by Wang et al.~\cite{wang20023}. Xu~\cite{xu2006thermal} used ANSYS based on the FEM to perform thermal simulation for a quad-core chip. COMSOL was used by Vaddina et al. ~\cite{vaddina2012thermal} to conduct thermal modeling and analysis of 3D stacked structures. With rapid advance in multi-core processors~\cite{AMD1}, localized power density has been significantly enhanced and it is desirable to attain high enough resolution in a thermal prediction to capture small-size hot spots. Although, DNS offers high thermal resolution and high accuracy, it is impractical to apply DNS at the chip architecture level due to their computationally intensive nature. }
\subsection{Thermal Circuit Model}
Lumped element thermal circuits have been widely used at the architecture level due to their simplicity and efficiency~\cite{wu2005joint}.
{For example,  HotSpot~\cite{HotSpot1} and 3D-ICE~\cite{sridhar20103d} are some of the popular thermal-circuit architecture level simulators. HotSpot has been applied by Glocker et al. ~\cite{glocker2013modeling} to study the temperature distribution in a 16-core system and by Florea et al.~\cite{florea2014enhancing} to provide thermal analysis for enhancing the Sniper multicore simulator. HotSpot and 3D-ICE have been integrated into the performance-power-thermal simulation toolchains, such as HotSniper ~\cite{pathania2018hotsniper}, HotGauge ~\cite{hankin2021hotgauge} and CoMeT ~\cite{siddhu2022comet}.   The efficiency of the thermal circuit model is however accomplished by using large lumped elements that are not able to adequately account for distributed heat transfer and may lead to an inaccurate prediction~\cite{fetis2006evaluation}. To improve the accuracy, the size of the RC elements needs to be reduced substantially, as done in the grid model of HotSpot~\cite{huang2007improved} that is then similar to the computationally intensive FDM.}
\subsection{Green’s Function Method} 
In the conventional Green’s function approach, the single Green’s function is calculated in response to a unit point heat source at the center of a large chip, assuming no boundary influence. The Green’s function is a spatial impulse response of the chip, and the thermal solution is constructed by superposition of the impulse responses to the point sources at different locations. It is thus inherently difficult to include boundary conditions (BCs) in thermal simulation of a finite domain~\cite{sultan2019survey,zhan2007high} or to use Green's function in situations where the power source is close to the edge of the chip~\cite{sultan2020fast}. It is also difficult to apply the conventional approach to transient thermal simulation~\cite{sultan2019survey,varshney2019nanotherm}. The method is however significantly more efficient than the FEM, FVM or FDM because it only considers a single layer where the power sources are generated. Various efforts have been made to overcome the aforementioned limitations by implementing different techniques with some significant revisions, which inevitably reduce computing efficiency. For example, the Green’s function is applied to correct the corner and edge effects based on the method of image for the adiabatic BC~\cite{ziabari2014power}. This is however not valid for different types of BCs. Using the power blurring method, together with the spatiotemporal pulse, transient thermal simulation becomes possible within the Green’s function framework~\cite{ziabari2014power}. These however still offer 2D temperature on the power dissipation layer. An effort has been made to include 3D temperature profile with multilayer Green’s functions~\cite{zhan2005high}; it is however limited to steady state simulation.

\subsection{Machine Learning Techniques}
Machine learning methods have recently been applied to investigate various aspects in semiconductor chips, including multi-core chips,~\cite{zhang2017machine,sridhar2011neural,qian2015support}. Among different machine learning techniques, the neural network methods have been the most common approaches for thermal simulation of semiconductor chips due to their simplicity and efficiency~\cite{zhang2017machine,sridhar2011neural}. Using neural networks, thermal data are needed to train each thermal cell (or neuron) in the domain to respond to heat excitations (heat sources and BCs) and neighboring neurons. Memory space and computational time needed in thermal simulation are thus determined by the selected number of neurons and the selected distance for close connections of the individual neurons. Although these approaches are considerably more efficient than the DNSs; they are in general limited to the thermal solution over a small number of nodes, instead of the detailed thermal profile of the chip. In addition, the approaches are based on data statistics without being guided by fundamental physical principles and thus in general leads to erroneous solution in case of extrapolation. 

\subsection{Physics-based Learning Method}
A different learning algorithm derived from a projection-based reduced order model has been investigated in the past based on POD~\cite{lumley1967structure,berkooz1993proper}. A tutorial for POD is included in ~\cite{weiss2019tutorial} that offers an intuitive and easy-to-follow presentation about the basic concepts and approach. The POD projects the thermal problem from a physical domain to a functional space to reduce the DoF. Instead of assuming its basis functions as done in many other projection-based approaches, such as  the Fourier series, Legendre polynomial, Bessel functions, etc., the basis functions (or POD modes) that constitute the POD space are generated (or trained) via thermal solution data for the problem of interest. The thermal data are collected from accurate DNSs for the trained POD modes to capture the essential thermal information induced by the parametric variations, including heating power and BCs. The POD offers an optimal set of basis functions that lead to a smallest least square (LS) error limited by the quality of the thermal data. To incorporate the physics principles of heat transfer in POD simulation, the dynamic heat transfer equation for the simulation domain is further projected onto the POD space. This rigorous procedure guides the POD thermal solution to comply with the dynamic heat transfer equation. Cheng’s group has applied the POD simulation methodology to develop efficient and accurate thermal simulation models for SOI and FinFET devices, FinFET  circuits at the gate level and interconnects~\cite{jia2022methodology}. 

The major advantage of the POD simulation methodology, guided by fundamental physics principle, is to reduce a large-dimension problem to an extremely small dimension with only a handful of modes while maintaining the accuracy comparable to and the resolution equivalent to DNS. A post process is however needed to reconstruct the dynamic temperature in the physical domain, which is the major computational bottleneck. Unlike the DNS where the entire simulation time and domain needs to be included in the simulation, the post processing calculations in the POD approach for realistic situations only need to be carried out in small high-temperature areas of the chip at each selected interval of time. This will significantly minimize the computing time and memory space needed in the post process.

\section{POD Thermal Simulation Methodology}

As described above, the POD based simulator, PODTherm-GP, is able to model a large-scale dynamic thermal problem using a small number of POD modes~\cite{jia2014thermal,jia2022methodology}. These modes are trained by the thermal solution data of the physical domain obtained from DNSs subjected to a range of parametric variations including the BCs and dynamic power map. The POD modes are thus specifically tailored to the training range of the BCs and power variations for the problem of interest and able to substantially reduce the DoF for the heat transfer problem. {As will be demonstrated in Sec.~\RNum{5}, PODTherm-GP is also able to offer an accurate prediction even beyond the training range if more modes are included because of the projection of the heat transfer equation onto the POD space. The projection onto the POD modes clearly incorporates the physical principle of heat transfer that allows the POD approach to predict the dynamic thermal evaluation in space adequately even in the extrapolation situation.}

\subsection{Generation of POD Modes}
The POD modes  are optimized by maximizing the mean square inner product of the thermal solution data with the modes over the entire domain~\cite{lumley1967structure,berkooz1993proper}, subjected to dynamic or static parametric variations. In our study, the thermal data in space is collected at each simulation time step in the DNS. This maximization process leads to an eigenvalue problem described by the Fredholm equation,
\begin{equation}
   \int_{   { \Omega }  } R( \vec{r},   \vec{r} '  )  \varphi (  \vec{r} ' ) d  \vec{r} '  =  \lambda   \varphi( \vec{r } ),  \label{Eq:1} 
\end{equation}
where $\lambda$ is the eigenvalue corresponding to the POD mode (i.e., eigenfunction $ \varphi$)  and $R( \vec{r},   \vec{r} '  )$ is a two-point correlation tensor given as
\begin{equation}
R( \vec{r},   \vec{r} '  )= \langle T(  \vec{r} , t)  \otimes T(  \vec{r}' , t)\rangle,\label{Eq:2}
\end{equation}
with $\otimes$ as the tensor operator and the angle brackets $\langle \rangle$ indicate the average over a number of thermal data sets. After solving the eigenvalue problem in~\eqref{Eq:1}, temperature can then be represented by 
\begin{equation}
    T( \vec{r}, t ) =  \sum_{i=1}^M  a_{i}(t)  \varphi _{i}( \vec{r} ),\label{Eq:3}
\end{equation}
where $a_{i}$  are the weighting coefficients for $\varphi _{i}$ and $M$ is the number of POD modes selected to reconstruct the temperature solution. If the eigenvalue of the data decreases rapidly for the higher modes, only a small  DoF (or $M$) is needed to reach an accurate thermal prediction.

\subsection{Projection of Heat Transfer Equation onto POD Space}

In order to predict the dynamic thermal distribution given in~\eqref{Eq:3} in a domain structure, the coefficients $a_{i}(t)$ need to be determined. To achieve this, the heat transfer equation is projected onto the POD space using the Galerkin projection,
\begin{equation}
\begin{aligned}
     \int_ \Omega (\varphi _{i}( \vec{r} ) \frac{\partial  \rho CT}{\partial t} +  \nabla    \varphi _{i}   \cdot  k  \nabla  T ) d \Omega   =  \int_ \Omega    \varphi _{i} ( \vec{r} )  P_{d} ( \vec{r}, t ) d \Omega \\
     -  \int_S \varphi _{i} ( \vec{r} )(-k  \nabla  T \cdot   \vec{n})dS, 
\end{aligned}\label{Eq:4}
\end{equation}
where $k$, $\rho$ and $C$ are the thermal conductivity, density and specific heat, respectively, $P_{d} ( \vec{r}, t )$ is the interior power density, $S$ is the boundary surface and $ \vec{n}$ is the outward normal vector on the boundary surface. With the selected POD modes, ~\eqref{Eq:4} can be rewritten as an $M$-dimensional ordinary differential equations (ODEs) for $a_{i}(t)$, 
\begin{equation}
     \sum_{i=1}^M  c_{i,j} \frac{da_{i}(t)}{dt} +\sum_{i=1}^M  g_{i,j} a_{i}(t) =  p_{j}, \: j = 1\: \textrm{to}\:M, \label{Eq:5}
\end{equation}
where $c_{i,j}$ are the elements of the thermal capacitance matrix in the POD space and they are defined as 
\begin{equation}
    c_{i,j} =  \int_ \Omega   \rho C  \varphi _{i} \varphi _{j} d \Omega. \label{Eq:6}
\end{equation}
$g_{i,j}$ and $ p_{j}$ are the elements of the thermal conductance matrix and power vector in the POD space, respectively. In ~\eqref{Eq:5}, $g_{i,j}$ and $ p_{j}$ may also vary with BCs.  In this work, two kinds of BCs, adiabatic and convective BCs, are applied to the boundary surfaces of the chip. For the adiabatic surface, the heat flux is zero, and $g_{i,j}$ and $p_{j}$ can be written as
\begin{equation}
    g_{i,j} =  \int_\Omega  k  \nabla \varphi _{i} \cdot  \nabla \varphi _{j}d \Omega, \:  p_{j}= \int_ \Omega  \varphi _{j} P_{d} ( \vec{r}, t ) d \Omega. \label{Eq:7}
\end{equation}
For the convective boundary, the heat flux on the surface is described by
\begin{equation}
  -k  \nabla  T \cdot   \vec{n} = -k \dfrac{\partial T}{\partial n} = h(T - T_{amb}), \label{Eq:8}
\end{equation}
where  $h$ is the heat transfer coefficient and $T_{amb}$ is ambient temperature. {
Using~\eqref{Eq:8} in~\eqref{Eq:4}, $g_{i,j}$ for the convective BC are given by}
\begin{equation}
    g_{i,j} =  \int_\Omega  k  \nabla \varphi _{i} \cdot  \nabla \varphi _{j}d \Omega + \int_ S  h \varphi _{i} \cdot \varphi _{j}dS , \label{Eq:9}
\end{equation}
and $p_{j}$ becomes 
\begin{equation}
    p_{j} = \int_ \Omega  \varphi _{j} P_{d} ( \vec{r}, t ) d \Omega + \int_ S  h \varphi _{j} T_{amb}dS. \label{Eq:10}
\end{equation}

Once the dynamic power consumption is obtained from gem5 and McPAT (described in Sec.~\RNum{4}), the interior power source strength in POD space given in~\eqref{Eq:7} and~\eqref{Eq:10} can be pre-evaluated. The BC of the substrate bottom is modeled by convective heat transfer given
{in~\eqref{Eq:8} } with a heat transfer coefficient determined by structure dimensions and material properties of heat spreader and heat sink with $T_{amb}$ = 45 °C. All other boundaries are adiabatic. The coefficients in~\eqref{Eq:5} are all pre-evaluated once the modes are determined. With $a_{i}(t)$ solved from~\eqref{Eq:5} in POD simulation, the temperature solution can be predicted from~\eqref{Eq:3}.

\section{Evaluation Methodology}

\begin{figure}[tb]
    \centering
    \includegraphics[width=0.5\textwidth]{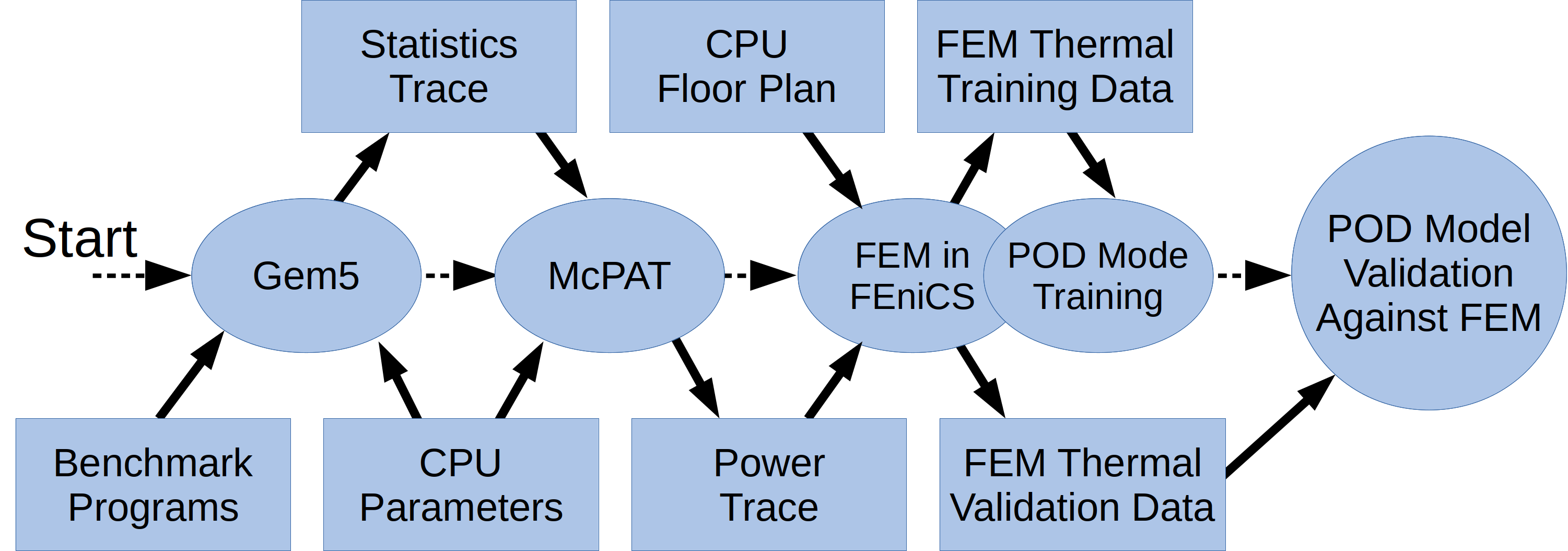}
    \caption{Workflow of the thermal simulation via PODTherm-GP for multi-core CPUs at the architecture level.}
    \label{fig:flow}
\end{figure}

The complete workflow of the thermal simulation via PODTherm-GP at the architecture level for multi-core CPUs is illustrated in Fig.~\ref{fig:flow}. An Intel Xeon E5- 2699v3 CPU with 18 cores, whose floorplan is shown in Fig.~\ref{fig:floorplan}~\cite{o1}, is selected in this work to demonstrate the effectiveness of PODTherm-GP. Using the benchmarks described below, the power trace in each unit of the CPU is generated from two open-source simulators, gem5 and McPAT, for FEniCS-FEM to perform thermal simulations. Gem5 was selected to perform cycle-level simulation of the CPU and to generate performance statistics of the CPU~\cite{GEM5_2011}. The configuration parameters used for gem5 to simulate the Intel Xeon E5-2699v3 CPU are shown in Table~\ref{tab:gem5conf}. These parameters are based on the specifications provided by Intel for the CPU and those used by Jongerius et al., in their work, where a similar Intel Xeon processor was modelled in gem5~\cite{ANALYTIC_2017}. Any unknown parameters are left as the default values configured by gem5.

\begin{table}[t]
    \caption{Gem5 Configuration Parameters for Intel XEON E5-2699v3}
    \centering
    \begin{tabular}{|c|c|}
        \hline
        \textbf{Parameter}                           & \textbf{Value}         \\ \hline
        Clock Speed                         & 2.3 GHz       \\ \hline
        Number of Cores                     & 18            \\ \hline
        L1$_i$ Size                         & 32 kB         \\ \hline
        L1$_d$ Size                         & 32 kB         \\ \hline
        L2 Size                             & 256 kB        \\ \hline
        L3 Size                             & 45 MB         \\ \hline
        Cache Line Size                     & 64            \\ \hline
        Memory Size                         & 16 GB         \\ \hline
        Memory Type                         & DDR3 1600 8x8 \\ \hline
        Issue Width                         & 8             \\ \hline
        ROB Capacity                        & 192           \\ \hline
        Number of Integer Units             & 4             \\ \hline
        Number of Floating Point Units      & 2             \\ \hline
        Integer Multiply Latency            & 3             \\ \hline
        Integer Division Latency            & 18            \\ \hline
        Floating Point Multiply Latency     & 5             \\ \hline
        Floating Point Division Latency     & 6             \\ \hline
        DRAM bandwidth                      & 68 GiB/s      \\ \hline
        DRAM Request to Response Latency    & 30ns          \\ \hline
    \end{tabular}
    \label{tab:gem5conf}
\end{table}

\begin{table}[htp]
 \caption{Selected SPLASH-2 Benchmarks}
    \centering
    \setlength{\tabcolsep}{2pt}
    \begin{tabular}{|C{2cm}|C{4cm}|C{2cm}|}
        \hline
        {\textbf{Benchmark}} & {\textbf{Description}} & {\textbf{Assigned Core Index}}\\ \hline
            Contiguous Ocean  & Simulates flow of large-scale oceanic currents using contiguous partitions & 2, 8, 14   \\ \hline
            Non-Contiguous Ocean  & Simulates flow of large-scale oceanic currents using non-contiguous partitions  & 3, 9, 15  \\\hline
            Radiosity  & Computes the distribution of light in a scene using the hierarchical diffuse radiosity method    & 1, 7, 13   \\\hline
            Water Spatial  & Simulates the flow of water in a space using a 3D spatial data structure          & 5, 11, 17   \\\hline
            Water N-Squared  & Simulates the flow of water in a space using a different algorithm than water-spatial   & 4, 10, 16  \\\hline
            Barnes  & Uses the Barnes-Hut method to simulate the interaction of a system of celestial bodies & 0, 6, 12  \\\hline
    \end{tabular}
    \label{tab:benches}
\end{table}

The SPLASH-2 benchmark suite was chosen as the simulated workload for gem5~\cite{SPLASH_1995}. SPLASH-2 is an open-source benchmark suites in C/C++, and it is widely used in the research field of computer architecture~\cite{noc_2016}. A subset of available benchmarks in SPLASH-2, described in Table~\ref{tab:benches}, was selected in this work. The assignment of these benchmarks for the generation of realistic dynamic power map is list in Table~\ref{tab:benches} using the index of the cores labeled in the Fig.~\ref{fig:floorplan} in order to mimic real-world scenarios. With these four benchmarks, two have variants: Ocean and Water. The variants of Ocean are contiguous and non-contiguous partitions. These variations change the way the algorithm partitions the simulated space. The variants of Water are spatial and N-squared. These variants change the algorithm used for the simulation. Any benchmark that was required to load data from an input file was modified so that the data provided with the benchmark was included in the source code. While this increased the size of the source code, the modification avoids requiring gem5 to read files from the disk. The CPU event traces output by gem5 include performance counters for different hardware components of the CPU, such as the usage of each functional unit, cache accesses, and many others. The simulator captures and stores as much information as possible about the CPU usage during each simulation time step. In this work, simulations are performed with a sampling interval of 4.35 $\mu s$ and the durations of 4.35 $ms$ and 6.09 $ms$ for the training and demonstration of PODTherm-GP, respectively.

The statistics traces output by gem5 for the selected benchmarks are parsed to create inputs for McPAT to simulate the power used by the CPU during operation. These input files are in the form of XML files that contain the required information for McPAT to operate. McPAT also requires architectural information, such as the number of functional units, the size of caches, and others. With the architectural information and performance statistics, McPAT estimates the area of each component, and calculates the dynamic and static power used by each component during each time step~\cite{MCPAT_2009}.  The power values output by McPAT are then parsed to create dynamic power map files. 

With the generated dynamic power map, FEniCS-FEM is used to perform thermal simulation for data collection to generate POD modes, as shown in Fig.~\ref{fig:flow}. The process for POD mode generation and POD model construction is further detailed in the next section. To validate the POD model, a separate FEniCS-FEM thermal simulation for the selected multi-core CPU is performed using a different dynamic power map. The POD results are then compared against the FEniCS-FEM simulation with the consistent numerical settings and dynamic power map.

\section{POD MODEL IMPLEMENTATION AND VALIDATION }

\subsection{Training of POD Modes}
\begin{figure}[tb]

    \centering
    \includegraphics[width=0.4\textwidth]{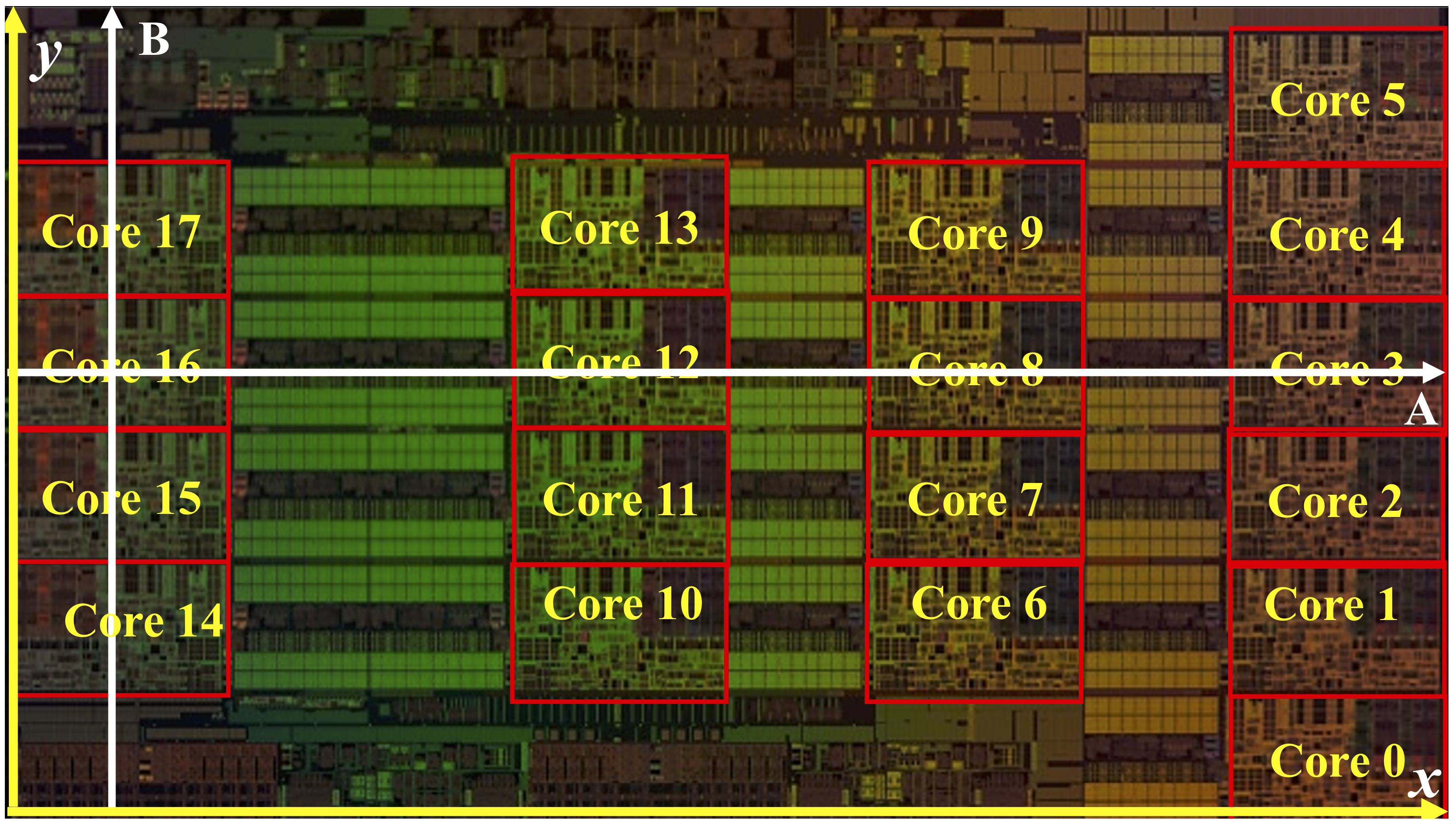}
    \caption{Floorplan of the Intel Xeon E5-2699v3 CPU~\cite{o1} {with an area of 31.0~$mm$ $\times$ 21.5~$ mm$}. A and B indicate the plotting paths for the demonstration.}
    \label{fig:floorplan}
\end{figure}

The simulation domain of the selected multi-core CPU, whose floorplan is shown in Fig.~\ref{fig:floorplan}~\cite{o1}, includes a top heating layer with a thickness of $55.8~um$ and a substrate with a thickness of $241.8~um$. The heating layer covers the top thickness where power is dissipated (such as the layers of devices and interconnects). Convective heat transfer described by a heat transfer coefficient given in~\eqref{Eq:8} is applied to the bottom of the substrate. The heat transfer coefficient is modeled by thermal resistances of the heat spreader, thermal interface material and heat sink calculated from the chip dimensions and material properties, similar to other studies~\cite{HotSpot1,huang2007improved}. The remaining boundary surfaces are assumed adiabatic. The dynamic power in each functional unit generated from gem5 and McPAT is averaged over 10k CPU cycles at 2.3 GHz. To investigate the effect of the training data quality on robustness of the POD model, two sets of thermal data are collected from FEniCS-FEM with meshes of 
$256\times256\times14$ and $1104\times620\times14$, or spatial resolutions of $0.121 \times 0.083 \times 0.023 ~mm^3$ and $0.028 \times 0.035 \times 0.023 ~mm^3$ in $x$, $y$ and $z$, respectively, using the same dynamic power map. The POD models built upon these 2 data sets with coarser and finer resolutions are referred to as POD Model-A and Model-B, respectively. It is noted that the coarser mesh used in this study is finer than those used in most of studies for chip-level thermal simulations, unless an extremely fine resolution is needed ~\cite{ziabari2014power,sultan2020fast}.

The POD modes and eigenvalues are determined from~\eqref{Eq:1} via the method of snapshots~\cite{sirovich1987turbulence,jia2022methodology} using the training data collected from FEniCS-FEM simulations. The POD modes are thus tailored to essential information embedded in the training data to account for variations of BCs and the dynamic power map. Each eigenvalue represents the mean squared temperature variations captured by its mode and therefore reveals the importance of the mode. The number of POD modes needed in~\eqref{Eq:3} for an accurate prediction can then be estimated from
\begin{equation}
     Err_{Theo} = \sqrt{  \frac{\sum_{i=M + 1}^{N_{s}} \lambda_{i}}{\sum_{i=1}^{N_{s}}   \lambda_{i}} }, \label{eq:LS_t}
\end{equation}
where $Err_{Theo}$ is the theoretical LS error and $N_{s}$ is the number of snapshots or sampled data sets. \eqref{eq:LS_t} offers an ideal LS error but the LS error resulting from PODTherm-GP is usually larger than $Err_{Theo}$ due to the limitations of numerical accuracy and computer precision.

\begin{figure}[tb]
 \centering
\includegraphics[width=0.4\textwidth]{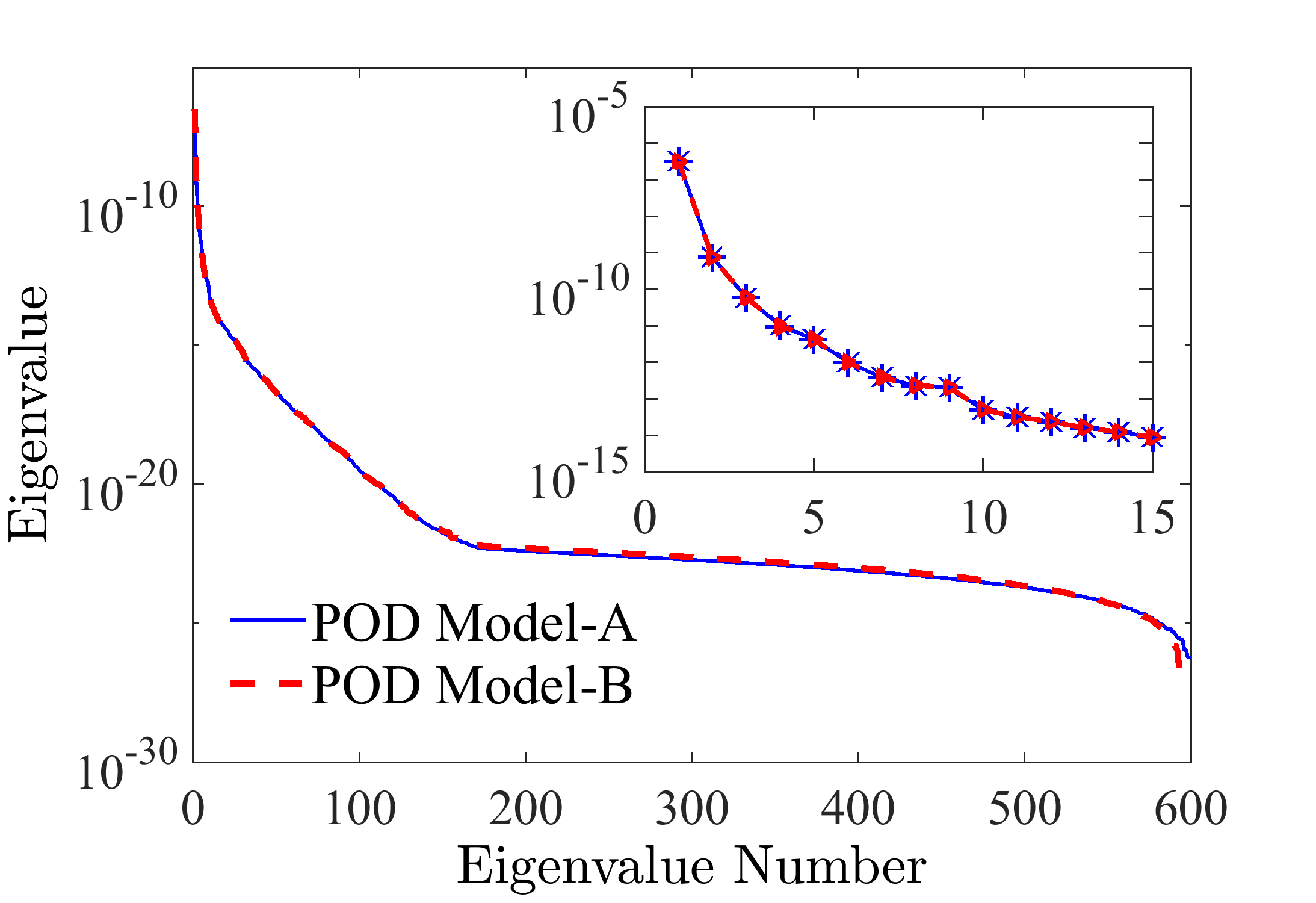} 
\caption{Eigenvalue spectrums of the thermal data for POD Model-A and POD Model-B.  A close-up inset shows the first 15 eigenvalues.}
\label{fig:eig}
\end{figure}

{ 
Fig.~\ref{fig:eig} describes the eigenvalues of the collected data for Model-A and Model-B in descending order. Since the resolution of the collected data sets are fine enough, the eigenvalues for these 2 POD models are nearly identical. The eigenvalue decreases significantly in the first several modes and shows a slower decreasing rate in the higher modes. It eventually becomes nearly flattened beyond 170 modes after a reduction by 16 orders of magnitude due to the limit of computer precision. The zoom-in spectrum in the inset shows that the eigenvalue drops more than two orders of magnitude from the first to the second mode and nearly four orders to the third mode.  Based on ~\eqref{eq:LS_t}, it is expected that either Model-A or Model-B is able to offer an accurate thermal prediction with 3 or more modes if the quality of the training data collected by the FEniCS-FEM is adequate. 
}

To construct a POD model, its model parameters in the projected ODEs in~\eqref{Eq:5} need to be pre-calculated using its POD modes~$\varphi _{i}( \vec{r} )$ of the training data, as defined in ~\eqref{Eq:6}-\eqref{Eq:10}. PODTherm-GP thus solves the ODEs for $\vec{a}(t)$ with a selected number of modes $M$, and a post process via ~\eqref{Eq:3} is then needed to calculate the dynamic thermal distribution in the CPU. Using the POD modes with two different resolutions, the following demonstrations focus on validating the accuracy and efficiency of the POD simulation technique and examining how the resolution impacts the quality of the training data and the generated modes and thus the model accuracy. A validation is also performed beyond the training range to examine the robustness of PODTherm-GP in the case of extrapolation.

\subsection{Demonstrations of the POD Simulation Methodology }
In the demonstrations, each of Model-A and Model-B is applied to thermal simulation of the selected multi-core CPU, compared to the FEniCS-FEM simulation with the spatial resolution identical to that of its POD modes. The training is carried out over a simulation time of 4.35 $ms$ for both Model-A and Model-B, and the POD models are validated against FEniCS-FEM for 4.35 $ms$ and 6.09 $ms$. In any comparison between simulations of a POD model and FEniCS-FEM, BCs and dynamic power map are identical. However, the dynamic power map used for the demonstration is different from that used for the collection of training data, as discussed in Sec.~\RNum{4}. 

\begin{figure}[tb]
 \centering
\includegraphics[width=0.49\textwidth]{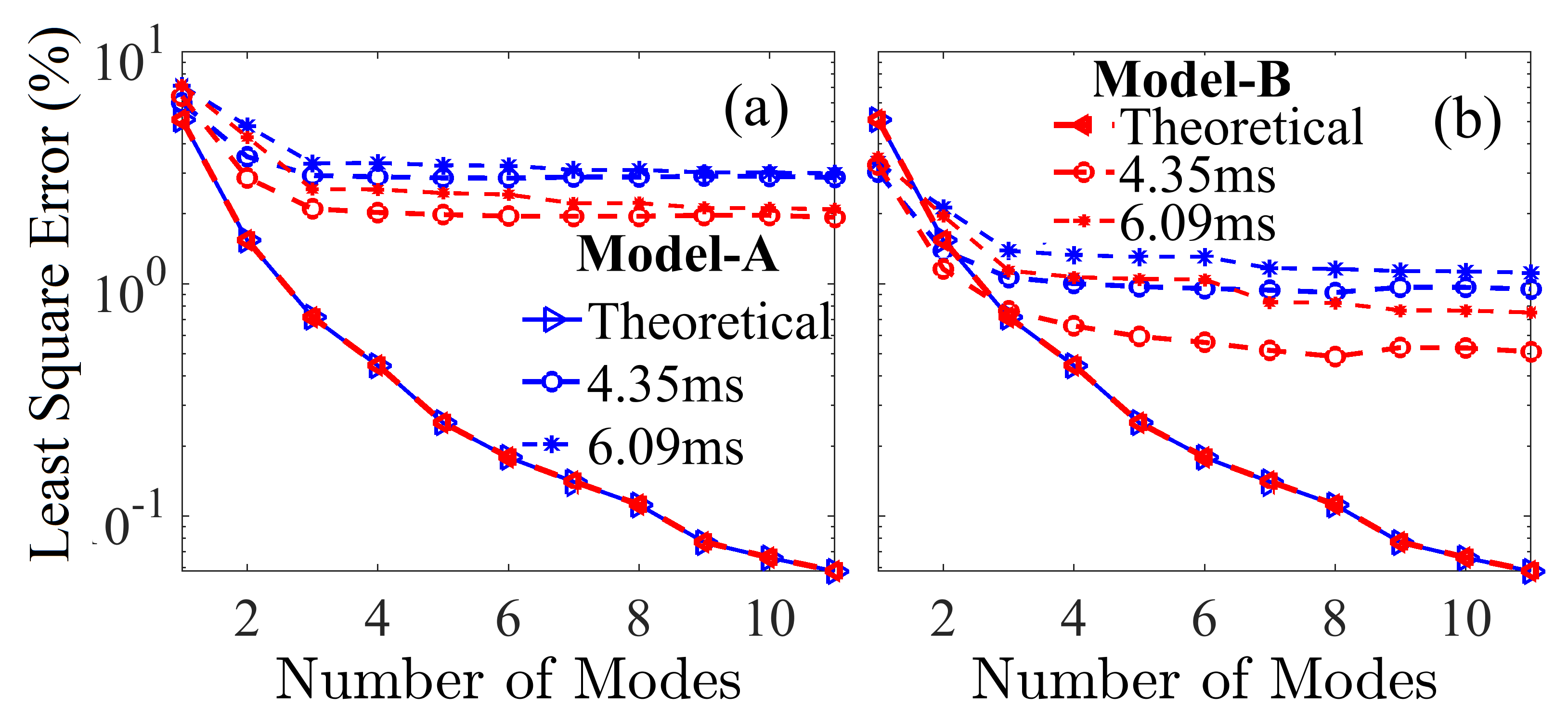} 
\caption{Least square errors for thermal simulations of the multi-core CPU in (a) the entire chip and (b) the heating layer.}
\label{fig:LS_error}
\end{figure}

Fig.~\ref{fig:LS_error} illustrates the LS errors of both POD Model-A and Model-B for the thermal simulations of the selected multi-core CPU, with respect to FEniCS-FEM simulations. The numerical LS error of the POD model is estimated over the entire simulation time from
\begin{equation}
     Err_{Num} = \sqrt{  \frac{\sum_{i=1}^{N_{t}}   \int_ \Omega    e_{i} ^{2}( \vec{r} ) d \Omega}{\sum_{i=1}^{N_{t}}   \int_ \Omega    (T_{i}( \vec{r} )-T_{amb}) ^{2} d \Omega} }, \label{Eq:12}
\end{equation}
where $T_{i}(\vec{r})$ and $e_{i}(\vec{r})$ are temperature given by FEniCS-FEM and the temperature  difference between the FEniCS-FEM and the POD model at $i$-th time step, respectively, and $N_{t}$ is the total number of time steps. 

As can be seen, the theoretical LS errors for Model-A and Model-B basically overlap due to the nearly identical eigenvalue spectrum shown in Fig.~\ref{fig:eig}. The theoretical LS error decreases rapidly from 5.1\% for one mode to 0.71\% with 3 modes. When using 3 modes, the numerical LS error shown in Fig.~\ref{fig:LS_error}(a) in the entire chip however reach 2.9\% for Model-A (coarser resolution) and 2.1\% Model-B (finer resolution) for the 4.35 $ms$ simulation. Beyond 3 modes, the error from the coarser mesh model remains nearly unchanged but from the finer mesh model continues decreasing slowly to 1.93\%. For the heating layer in Fig.~\ref{fig:LS_error}(b), the LS error for the 4.35 $ms$ simulation with 3 modes using Model-A or Model-B becomes 1.1\% or 0.76\%, respectively. The error from Model-A continues decreasing to 0.91\% with  8 modes while the error from Model-B decreases to 0.49\% with 8 modes and stays near 0.51\% beyond 8 modes. The errors in the entire chip and the heating layer are clearly improved when the POD modes are generated from the finer mesh training data.

For the 6.09 $ms$ case with simulation time beyond the training range, as expected, the LS error of the POD model increases. However, Figs.~\ref{fig:LS_error}(a) and (b) clearly show that, when more modes are included in this extrapolation case, the LS error of the 6.09 $ms$ simulation declining toward the error of the 4.35 $ms$ case. For example, when using 3-11 modes, the LS error of Model-B for the entire chip remains near 1.93\%-2.1\% for the 4.35 $ms$ case, and its error for the 6.09 $ms$ case is equal to 2.5\% with 3 modes and reduces to 2.2\% and 2.09\% with 7 and 11 modes, respectively. It is observed as well for Model-A in Fig.~\ref{fig:LS_error}(a) that the 6.09 $ms$ curve declines with more modes and these 2 LS-error curves nearly merge with 10 or more modes. The similar phenomenon is also shown in Fig.~\ref{fig:LS_error}(b) for the heating layer, when comparing the LS errors induced by these two POD models using more modes. 

As demonstrated in the 4.35 $ms$ case, for Model-A with 3 modes compare to FEniCS-FEM, a reduction in the DoF over 5 orders of magnitude ($256\times256\times14$ vs. $3$) with a LS error of 2.9\% for the entire chip or 1.1\% in the heating layer can be achieved.  For Model-B with 3 modes, a reduction of DoF more than 6 orders of magnitude ($1104\times620\times14$ vs. $3$) with an LS error of 2.1\% or 0.76\% for the entire chip or the heating layer, respective, can be reached. In this 4.35 $ms$ case, if 7 modes are used, an LS error as small as 1.95\% or 0.52\% can be accomplished over the entire chip or in the heating layer, respectively, and the reduction in the DoF is still as large as 6 orders.

To understand the insight into the LS errors presented in Fig.~\ref{fig:LS_error}, how the dynamic thermal profile influenced by the accuracy of the POD modes resulting from the quality of the training data is examined below in detail.  

\begin{figure}[t]
 \centering
\includegraphics[width=0.455\textwidth]{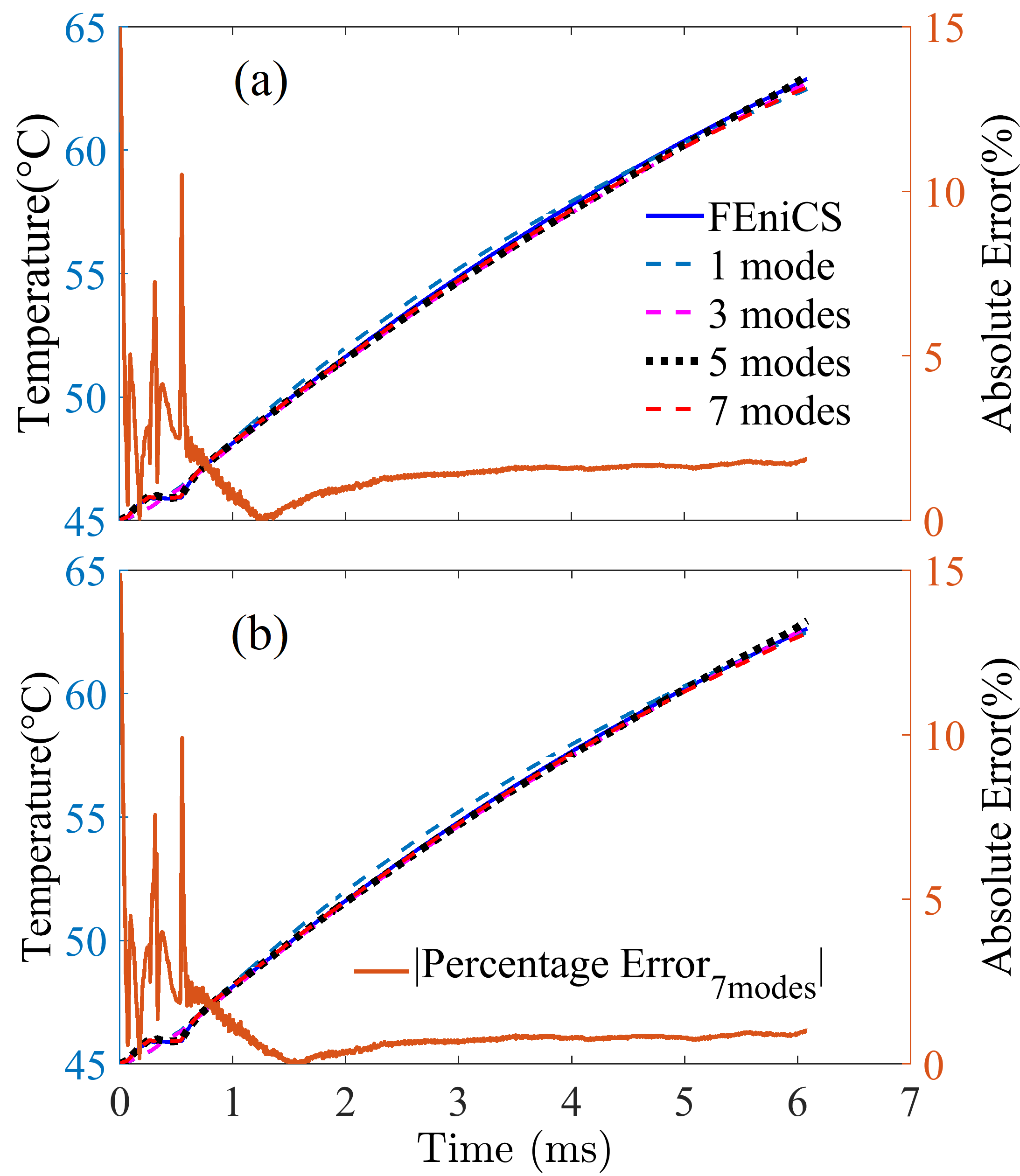} 
\caption{Comparison of temperature evolution at (2.3 $mm$, 11.8 $mm$), the intersection of Paths A and B shown in Fig.~\ref{fig:floorplan}, between FEniCS-FEM and (a) Model-A and (b) Model-B. The absolute percentage error for each POD model using 7 modes is also included.}
\label{fig:T_evolution}
\end{figure}

The temperature evolution at the intersection of Paths A and B predicted by POD Model-A and Model-B is compared against the FEniCS-FEM result in Figs.~\ref{fig:T_evolution}(a) and (b), respectively, over a simulation time of 6.09 $ms$. At this particular location, both POD-Galerkin models with 3 or more modes offer a very good agreement with FEniCS-FEM even beyond the training time (4.35 $ms$). When using 7 modes, the absolute percent errors for $t > 2.2~ms$ become nearly flattened and stay near or below 1.62\% from Model-A and 0.85\% from Model-B, where the finer-resolution model leads to a higher accuracy. Because $T_{amb}$ is used as the initial chip temperature, the percentage error w.r.t. $T_{amb}$ looks evidently large for $t < 0.55~ms$ even though the absolute error is less than $0.1^\circ$C at $t\approx 0.55~ms$ and much less than $0.01^\circ$C near $t = 0^+~ms$.

\begin{figure}[tb]
 \centering
\includegraphics[width=0.45\textwidth]{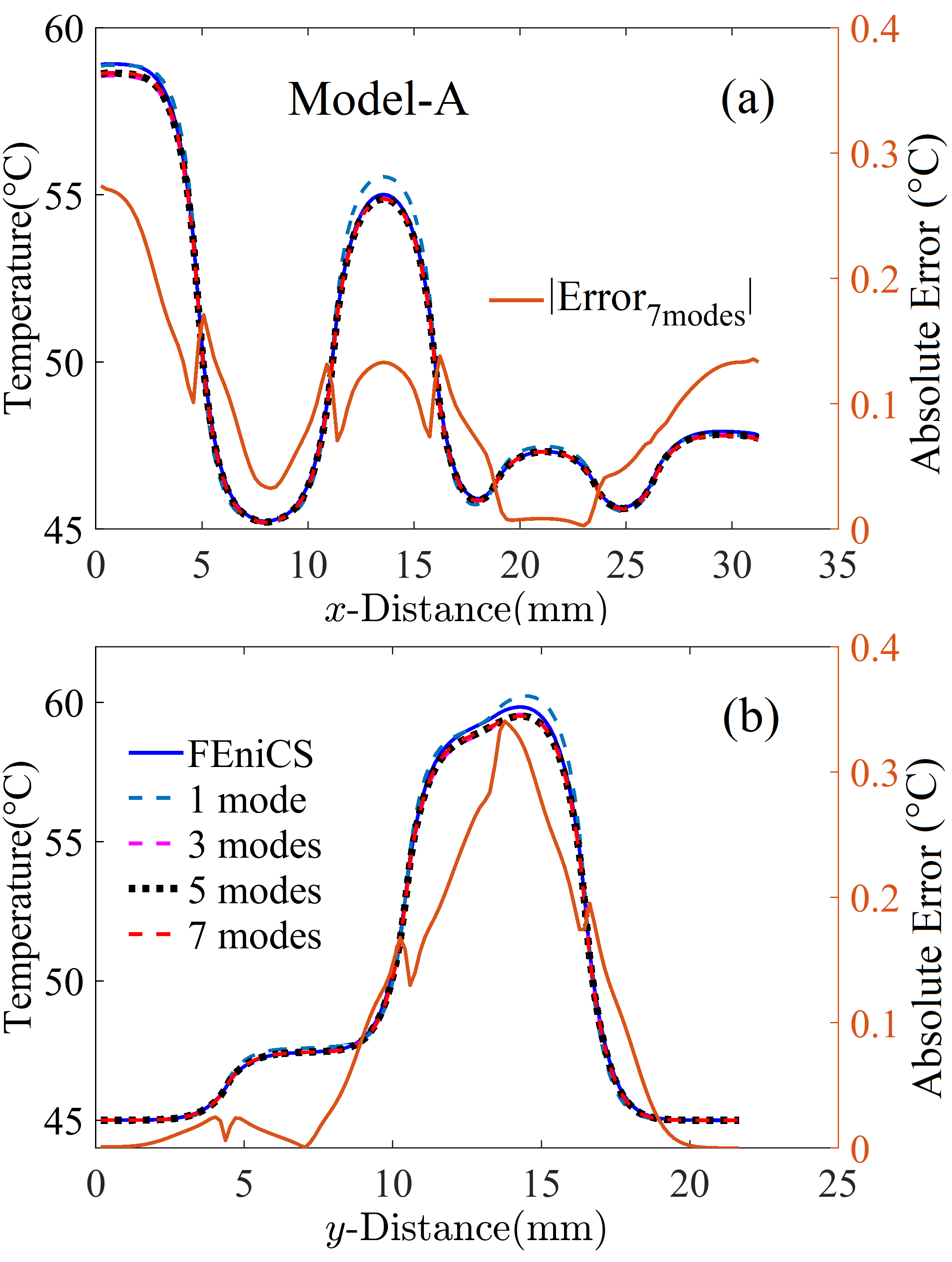} 
\caption{Temperature distribution derived from POD Model-A along (a) Path A and (b) Path B at 4.35~$ms$, compared to FEniCS-FEM. The absolute error using 7 modes is also included.}
\label{fig:Profile_A_4}
\end{figure}

\begin{figure}[tb]
 \centering
\includegraphics[width=0.45\textwidth]{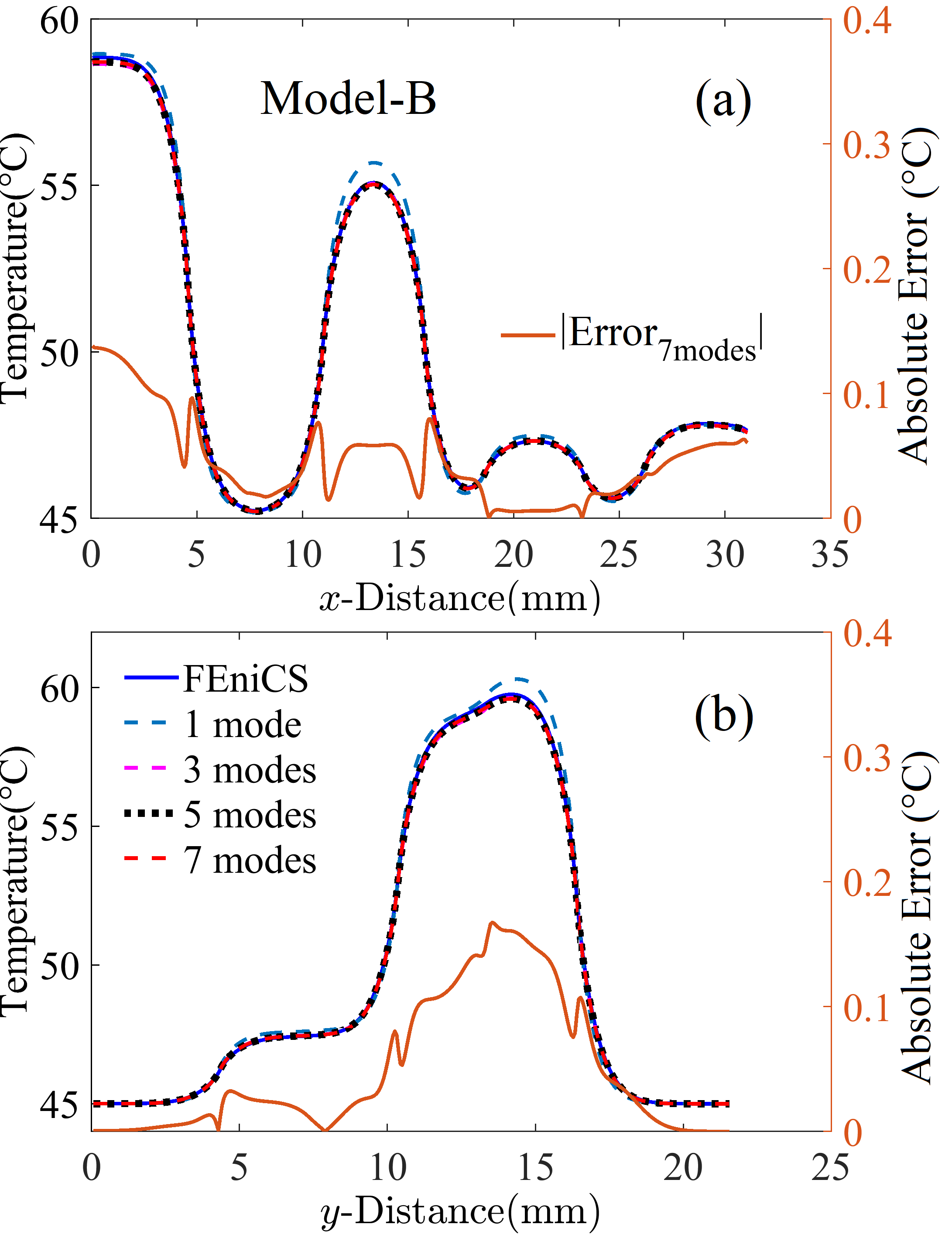} 
\caption{Temperature distribution derived from POD Model-B along (a) Path A and (b) Path B at 4.35~$ms$, compared to FEniCS-FEM.  The absolute error using 7 modes is also included.}
\label{fig:Profile_B_4}
\end{figure}

The temperature profiles along Paths A and B are shown in Fig.~\ref{fig:Profile_A_4} for Model-A and in Fig.~\ref{fig:Profile_B_4} for Model B. Based on the CPU floorplan in Fig.~\ref{fig:floorplan}, Figs.~\ref{fig:Profile_A_4} and ~\ref{fig:Profile_B_4} show that temperatures in Cores 12, 16 and 17 appear to be  higher than other functional units of the multi-core CPU resulting from the core assignment based on the selected benchmarks and their computational intensities. In general, POD Model-A with 3 modes offers an accurate thermal prediction along Paths A and B in the chip, as presented in Fig.~\ref{fig:Profile_A_4}. However, in the high temperature regions, Cores 16 and 17, there is a 2\%-3\% deviation from the FEniCS-FEM result. When a finer mesh is used in FEniCS-FEM to collect the training data to generate the POD modes, Model-B with 3 or more modes offers an excellent thermal prediction along both Paths A and B, as shown in Fig.~\ref{fig:Profile_B_4}. As also illustrated, the error derived from Model-B (a finer mesh in Fig.~\ref{fig:Profile_B_4}) is 50\% less than that from Model-A (a coarser mesh in Fig.~\ref{fig:Profile_A_4}) in nearly all locations along Paths A and B.

\begin{figure*}[t]
 \centering
\includegraphics[width=0.95\textwidth]{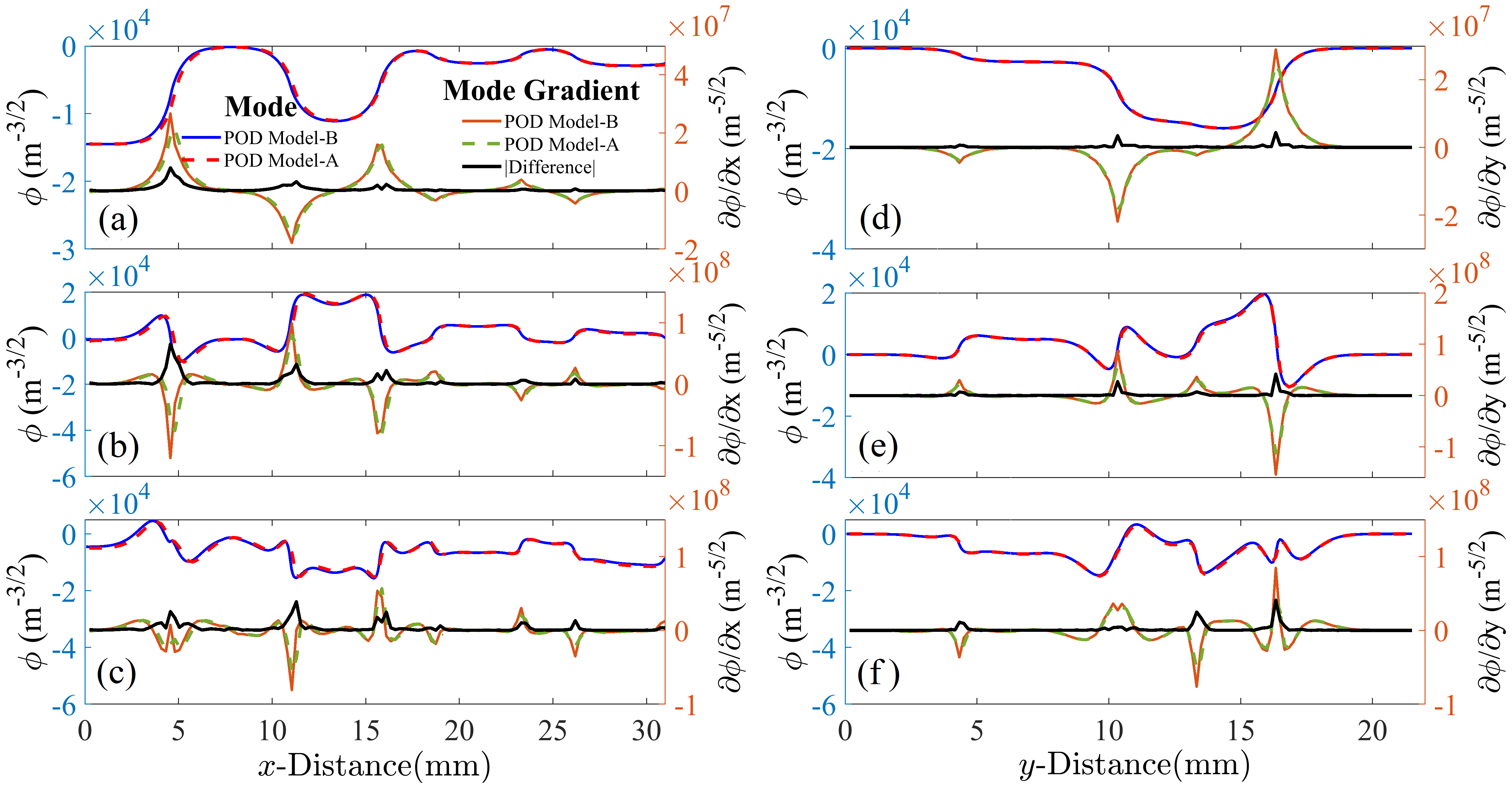} 
\caption{POD modes and their gradients along Path A for the (a) first, (b) second, (c) third modes; and along Path B for the (d) first, (e) second, (f) third modes.}
\label{fig:POD_mode}
\end{figure*}

To understand how the mesh size impacts the POD model accuracy, the profiles of POD modes and gradients of the modes for Model-A and Model-B are examined in Fig.~\ref{fig:POD_mode}. The difference of the mode gradients between Model-A and Model-B along Paths A and B are also included. Because the eigenvalue spectrums for these 2 POD models are nearly identical, as given in Fig.~\ref{fig:eig}, their modes are very close. However, a close look at the slopes of the modes in each direction reveals the hidden problem of the model built upon the coarser mesh. As shown in Fig.~\ref{fig:POD_mode}, the gradient differences of the POD modes between Model-A and Model-B at some locations with larger thermal gradients is relatively large. For example, as shown in Figs.~\ref{fig:POD_mode}(a)-\ref{fig:POD_mode}(c), a slope difference of the POD modes in $x$ around 25\%, 75\% or 45\% is observed in Modes 1, 2 or 3, respectively, near $x = 4.9~mm$ where the highest thermal gradient in $x$ is observed in Figs.~\ref{fig:Profile_A_4}(a) and ~\ref{fig:Profile_B_4}(a). Evident slope differences of the modes in $x$ are also observed in other large slope locations near $x = 12.4~mm$ and $16.15~mm$. Along Path B, the slope difference in $y$ shown in Figs.~\ref{fig:POD_mode}(d)-\ref{fig:POD_mode}(f) is smaller but still evident for the first 2 or 3 modes at $y =10.4~mm$ and $17.9~mm$ but the difference at $13.4~mm$ is observed only in the third mode.  

Accuracy of the POD model in~\eqref{Eq:5} is dictated by the quality of the coefficients evaluated in~\eqref{Eq:6}-\eqref{Eq:10}. The thermal conductance elements $  g_{i,j}$ in the POD space results from the projection of the heat flux ($-k  \nabla  T$) onto the POD modes and thus strongly depends on gradients of the modes, as defined in~\eqref{Eq:7} and ~\eqref{Eq:9}. The accuracy of the POD model is thus significantly influenced by the gradients of the generated POD modes that are directly impacted by the quality of the training data. This explains why Model-B built upon the thermal data collected from the FEniCS-FEM simulation with a finer mesh gives rise to a considerably smaller LS error than Model-A, as discussed above in Fig.~\ref{fig:LS_error}.

\begin{figure}[tb]
 \centering
\includegraphics[width=0.4\textwidth]{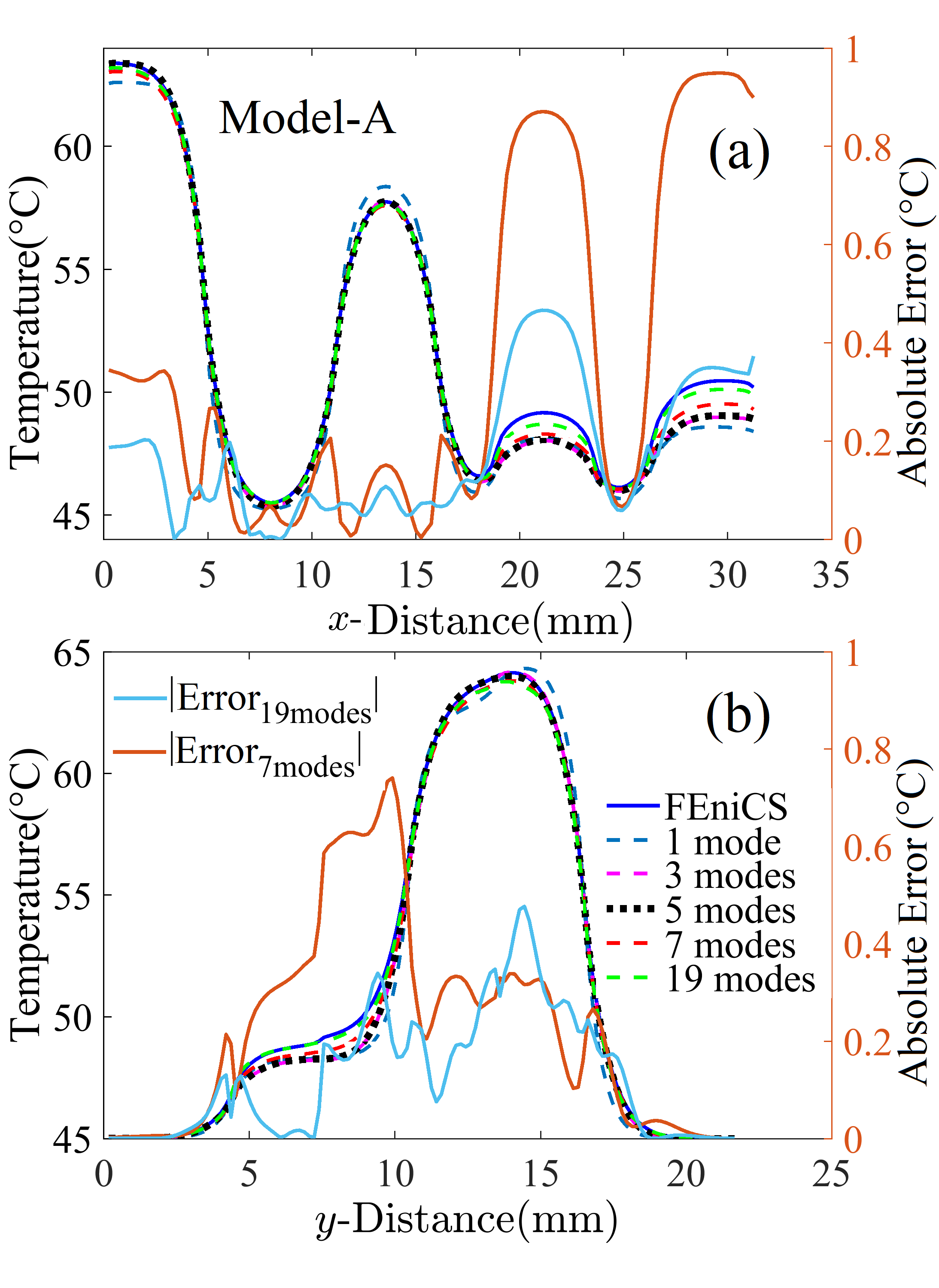} 
\caption{Temperature distribution derived from POD Model-A along (a) Path A and (b) Path B at 6.09~$ms$, compared to FEniCS-FEM.  The absolute errors using 7 and 19 modes are also included.}
\label{fig:Profile_A_6}
\end{figure}

\begin{figure}[tb]
 \centering
\includegraphics[width=0.4\textwidth]{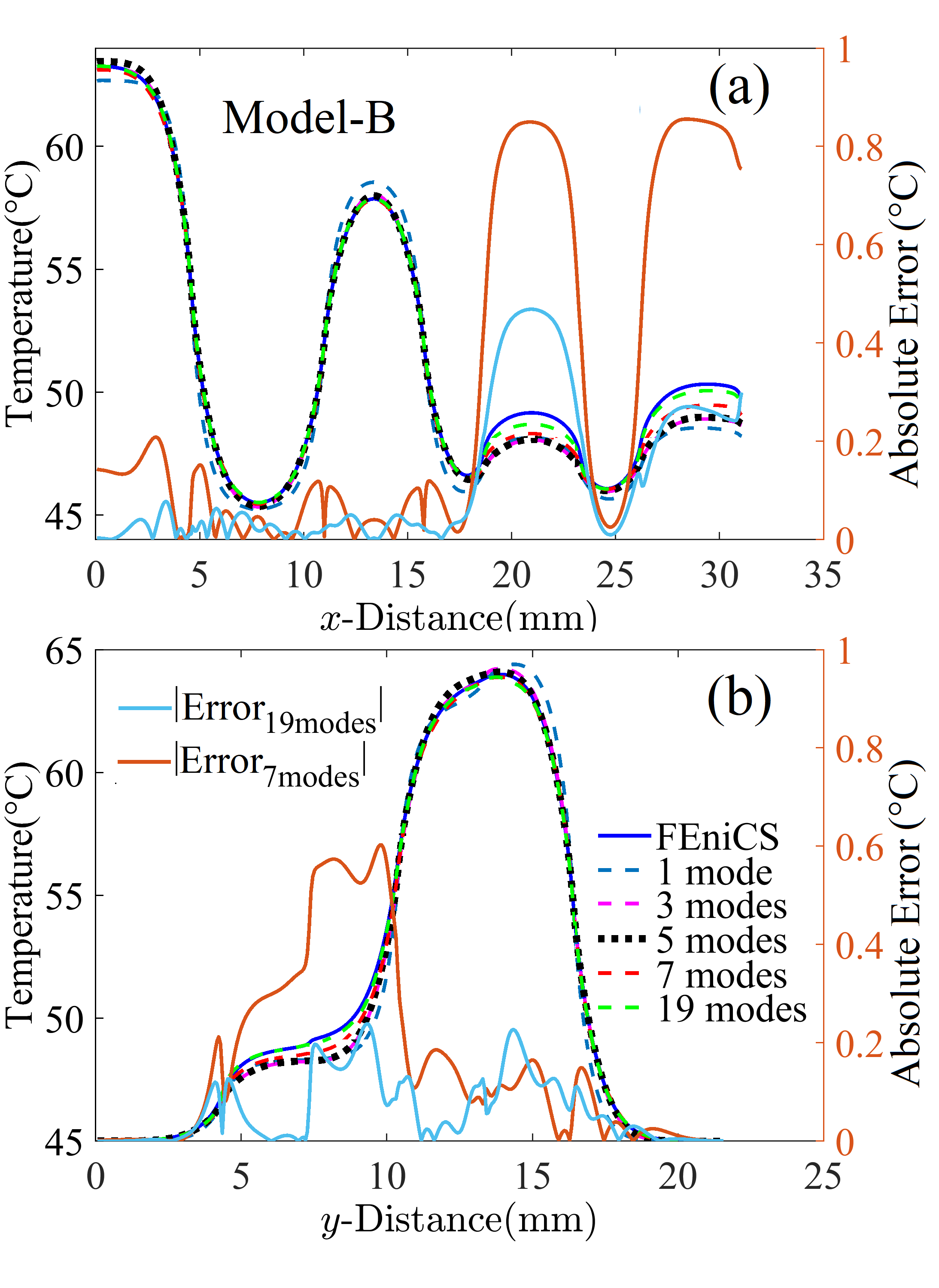} 
\caption{Temperature distribution derived from POD Model-B along (a) Path A and (b) Path B at 6.09~$ms$, compared to FEniCS-FEM.   The absolute errors using 7 and 19 modes are also included.}
\label{fig:Profile_B_6}
\end{figure}

To further examine the robustness of PODTherm-GP in the extrapolation case, the thermal profiles at 6.09~$ms$ predicted by Model-A and Model-B are illustrated in Figs.~\ref{fig:Profile_A_6} and \ref{fig:Profile_B_6}, respectively, compared to FEniCS-FEM results. For the case with a coarser mesh in Fig.~\ref{fig:Profile_A_6}, the thermal profiles along Paths A and B from POD Model-A with 3 or 5 modes agree reasonably well with those from FEniCS-FEM, except for the location near $x < 3~mm, 19~mm < x < 23~mm$, and $x > 27~mm$ in Fig.~\ref{fig:Profile_A_6}(a) and $5~mm < y < 10~mm$ in Fig.~\ref{fig:Profile_A_6}(b). Fig.~\ref{fig:LS_error}(b) illustrates that, with 7 or more modes, Model-A starts improving its LS error for the simulation beyond the training period and a better agreement with FEniCS-FEM is observed in Figs.~\ref{fig:Profile_A_6}(a) and (b). Because of considerably higher power applied to Cores 3 and 8 and a longer simulation time than the training conditions, it requires 19 modes to significantly improve the prediction from Model-A.  The absolute error included in Fig.~\ref{fig:Profile_A_6} also reveals a significant improvement when using 19 modes, compared to 7 modes although an error as small as 1.5\%-2\% still remains near $13~mm < x < 15~mm$ in Core 17 (similar to Fig.~\ref{fig:Profile_A_4} (b)). For the finer mesh case in Fig.~\ref{fig:Profile_B_6},  a similar improvement is obtained for the Model-B simulation beyond the training period when more modes are used, and a considerably better prediction is observed for $x < 3~mm$ and $13~mm < y < 15~mm$, where an excellent agreement with FEniCS-FEM is observed.

Results shown in Figs.~\ref{fig:LS_error},~\ref{fig:Profile_A_6} and ~\ref{fig:Profile_B_6} demonstrate that, even with core power and simulation time beyond the training conditions, PODTherm-GP offers a prediction with a reasonably good accuracy with just 7 modes and more modes are needed to reach a very accurate prediction. Such an extrapolation capability stems from the Galerkin projection of the heat transfer equation. Unlike most machine learning methods that usually lead to an unpredictable solution when simulation settings are outside the bounds of the training conditions, the Galerkin projection enforces the prediction to comply with the physical principles embedded in ~\eqref{Eq:5}. These results also confirm the significant influence of data quality (resulting from the mesh size in this case) on the prediction accuracy of PODTherm-GP in the extrapolation case. 

\section{discussion}
Many interesting and encouraging findings are observed in simulations of POD Model-A and Model-B, compared against FEniCS-FEM simulations within and beyond the training time.  The LS errors in the entire chip and the heating layer are clearly improved when the POD modes are generated from the finer-mesh training data. Within the training time using just 3 modes, as shown in Fig. ~\ref{fig:LS_error} Model-A and Model-B reach the LS error near 2.9\% and 2.1\%, respectively, for the entire chip. The finer mesh model (Model-B) can achieve an error near 1.95\% with 7 modes. If the temperature in the heating layer is the only concern, Model-A and Model-B with 3 modes offer an LS error of 1.1\% and 0.76\%, and Model-B reaches an error as low as 0.52\% with 7 modes.  In the case of extrapolation with $6.09~ms$ simulation time, Fig.~\ref{fig:LS_error} shows that the LS error of the POD models increases, as expected; however, by incorporating more modes in the simulation, the LS error can be brough down closer to that for the interpolation case. This is observed in Figs.~\ref{fig:LS_error}, ~\ref{fig:Profile_A_6} and ~\ref{fig:Profile_B_6}, as well as in Table~\ref{tab:Tab3}. The LS error in Table~\ref{tab:Tab3} is estimated with respect to the result from the finer-mesh FEniCS simulation.

\begin{table*}[h]
  \caption{Least Square Error w.r.t. Finer-mesh FEniCS-FEM Simulation }
  \centering
  \label{tab:Tab3}
  \begin{tabular}{|c|c|C{0.5cm}|C{0.5cm}|C{0.5cm}|C{0.5cm}|C{0.5cm}|c|C{0.5cm}|C{0.5cm}|C{0.5cm}|C{0.5cm}|C{0.5cm}|}
  \hline
  \textbf{Case} &\multicolumn{6}{c|}{\textbf{4.35 $ms$}} &\multicolumn{6}{c|}{\textbf{6.09 $ms$}}\\
  \hline
  \multirow{2}{*}{Simulator} &FEniCS &\multicolumn{5}{c|}{POD Model-B (No. modes)} &FEniCS&\multicolumn{5}{c|}{POD Model-B (No. modes)}\\
    \cline{3-7}
    \cline{9-13}
    &Coarse &3 &5 &7 &11 &13 &Coarse &3 &5 &7 &11 &13\\
    \hline
    Heating layer (\%) &0.80 & 0.76 &0.59 &0.52 &0.51 &0.51 &0.72 &1.14 &1.05 &0.83 &0.75 &0.61\\
    \hline
     Entire chip (\%) &1.63 & 2.10 &1.98 &1.95 &1.93 &1.93 &1.51 &2.55 &2.46 &2.22 &2.09 &1.86\\
     \hline
\end{tabular}
\end{table*}

It is shown that the coarser mesh in the FEniCS-FEM simulation for POD Model-A is fine enough to produce reasonably good-quality data that thus leads to nearly identical eigenvalues and POD mode profiles of the first 3 modes to those for Model-B, as shown in Figs.~\ref{fig:eig} and~\ref{fig:POD_mode}. However, the coarser-mesh FEM simulation is not able to capture accurate gradients in some interfaces between functional units with a large temperature difference, and the training data lead to inadequate quality of the POD modes. Apparently, as presented in Fig.~\ref{fig:POD_mode}, this is the key issue that leads a larger LS error from Model-A than from Model-B. This is also the reason for an evident deviation observed between the coarser-mesh and finer-mesh FEM thermal simulations of the CPU, as shown in Table~\ref{tab:Tab3}. An LS error of the coarser-mesh FEM, compared to the finer-mesh FEM, as high as 0.8\% and 1.63\% is observed for the heating layer and entire chip, respectively. 

The error of the Model-B within and beyond the training period is also included in Table~\ref{tab:Tab3}. As more modes are included, the LS error for Model-B in both $4.35~ms$ and $6.09~ms$ cases declines, and also the error for the $6.09~ms$ case moves closer to that for the $4.35~ms$ case. In the entire chip, Model-B leads to an LS error slightly greater than the coarser-mesh FEM simulation. However, in the heating layer, Model-B with just 3 modes leads to a smaller error than FEniCS-FEM with a coarser mesh, and it reaches an error as low as 0.51\% with 13 modes, compared to 0.8\% error induced by the coarser mesh FEM. For the $6.09~ms$ case, use of 13 modes in Model-B reduces the LS error to 0.61\% that is smaller than 0.72\% for the coarser FEM simulation. As mentioned previously, the coarser mesh size used in our study is actually finer than those generally used in architecture-level thermal simulations ~\cite{ziabari2014power,sultan2020fast}. Even in the extrapolation case beyond the training time, if more modes are used, Model-B is actually more accurate than the FEM simulations with mesh sizes finer than what are usually used in most studies on the chip-level thermal simulations. 

As estimated in Sec.~\RNum{5}, the reduction in the DoF could be 5 to 6 orders of magnitude. To estimate the computational efficiency, the computational time for the thermal simulation of the selected multi-core CPU is listed in Table~\ref{tab:Tab4} for each approach. All experiments were conducted on the Dell Precision Tower T7910, with two Intel Xeon E5-2697A v4 CPUs, 512G memory, and Ubuntu 20.04 as its OS. As to PODTherm-GP, the computational time consists of the times for solving $a_{i}(t)$ in~\eqref{Eq:5} (denoted as ODEs in Table~\ref{tab:Tab4}) and the post process to recover temperature in physical space using ~\eqref{Eq:3}. It turns out the post process to restore the dynamic temperature in physical space for every time step requires considerably more effort than that solving $a_{i}(t)$ from the ODEs because of the high spatial resolution in the POD modes. However, the post processing calculation is very flexible. As previously mentioned, realistically, only temperature in some units, where computational workload is heavy, is needed. This will significantly reduce the computing time and memory space needed in the post process. Post1 and Post2 denote the post processes for restoring the dynamic thermal profiles in the heating layer and the entire CPU, respectively. Model-A with 3 modes leads a computational speedup of more than 19,000 and 5,500 times (i.e., 2,2880/(0.082+1.098) and 2,2880/(0.082+4.034)) in the heating layer and entire chip, respective, compared to FEniCS-FEM with the same resolution as Model-A. When using 5 modes, the speed improvement over the FEniCS-FEM is still as high as 12,900 and 3,200 times, respectively. For the POD Model-B simulation compared to the finer mesh FEniCS-FEM simulation, the saving in computational time is 3.5 - 4 times higher than what Model-A offers.

When comparing Model-B (the finer mesh) to FEniCS-FEM with the coarser mesh, the saving in computing time in the entire chip is more than 2,100 and 1,200 times (i.e., 2,2880/(10.604+0.081) and 2,2880/(18.0+0.101)) using 3 and 5 modes, respectively. When considering only the heating layer, Model-B offers a speedup over 7,700 and 5,000 times using 3 and 5 modes, respectively, compared to the FEniCS-FEM with the coarser mesh, and yet Model-B in the heating layer is more accurate with just 3 modes (see Table~\ref{tab:Tab3}).

\begin{table*}[t]
  \caption{Computational Time for Thermal Simulations of the CPU }
  \centering
  \label{tab:Tab4}
  \begin{tabular}{|c|c|c|c|c|c|c|c|c|c|c|}
  \hline
 \textbf{ Mesh} &\multicolumn{10}{c|}{\textbf{Computational Time ($s$)}} \\
  \hline
   	\multirow{5}{*}{$256 \times 256 \times 14$ }&\multirow{4}{*}{POD Model-A} &\multicolumn{9}{c|}{No. of POD modes} \\
  \cline{3-11}
   & &\multicolumn{3}{|c|}{1}&\multicolumn{3}{|c|}{3}&\multicolumn{3}{|c|}{5}\\
   \cline{3-11}
  & &ODEs &Post1 &Post2 &ODEs &Post1 &Post2&ODEs &Post1 &Post2\\
  \cline{3-11}
  & &0.075 &0.572 &1.795 &0.082 &1.098 &4.034 &0.089 &1.677 &7.011\\
  \cline{2-11}
  &FEniCS-FEM &\multicolumn{9}{|c|}{$2.288 \times 10^4 $}\\
    \hline
    \multirow{5}{*}{$1104 \times 620 \times 14$ }&\multirow{4}{*}{POD Model-B} &\multicolumn{9}{c|}{No. of POD modes} \\
  \cline{3-11}
   & &\multicolumn{3}{|c|}{1}&\multicolumn{3}{|c|}{3}&\multicolumn{3}{|c|}{5}\\
   \cline{3-11}
  & &ODEs &Post1 &Post2 &ODEs &Post1 &Post2&ODEs &Post1 &Post2\\
  \cline{3-11}
  & &0.079 & 1.474 &4.688 &0.081 & 2.885 &10.604 &0.101 &4.406 &18.400\\
  \cline{2-11}
  &FEniCS-FEM &\multicolumn{9}{|c|}{ $2.247 \times 10^5 $}\\
  \hline

\end{tabular}
\end{table*}

It is worthwhile to mention that in realistic dynamic thermal management, it is not necessary to have the information on the temperature profile at every instant in time. Thus, the computational time needed for PODTherm-GP to restore the temperature in~\eqref{Eq:3} for every tens of time steps would be on the same order as that needed to solve the ODEs in~\eqref{Eq:5}. Unlike the DNS where temperature in the entire temporal and spatial domains needs to be included in the simulation, the computational speedup for PODTherm-GP will be at least one more order faster than what was discussed above if only temperature at a few instants over a certain period is needed in higher temperature areas. 

This finding of this investigation strongly suggests that a POD model could offer a more accurate thermal prediction than the DNS with a fine mesh if the POD model is built upon a finer mesh DNS.  Although the training time will be longer, the POD model will then outperform the DNS in terms of the accuracy with a computational speedup by at least 3 orders of magnitude. The saving in computational time for the POD model is at least 4 orders of magnitude in the realistic applications of dynamic thermal management since only temperature in a small fraction of the simulation time span and spatial domain is needed. 

Similar to other learning methods for complex problems, when applying PODTherm-GP to thermal simulation of a CPU or GPU with a large number of cores, the POD-Galerkin methodology would suffer from intensive computational effort to collect data for POD mode training. To resolve the training issue for a processor with hundreds or thousands of cores, domain decomposition can be applied to partition the processor into many smaller subdomains and each subdomain is trained separately with much less effort. Such an approach, together with different training strategies, will be investigated in the near future.
\section{Conclusion}
PODTherm-GP, an architecture-level thermal simulation methodology, has been investigated for an 18-core CPU, Intel Xeon E5-2699v3, in terms of its accuracy, efficiency and robustness. The methodology is developed based on the POD, together with the Galerkin projection of the heat transfer equation onto the POD modes that are trained by thermal data collected from accurate FEniCS-FEM simulation. The dynamic power maps implemented in the FEniCS-FEM thermal simulations of the selected CPU were generated by gem5 and McPAT, together with the SPLASH-2 benchmark suite. It has been shown that the quality of the training thermal data, especially the accuracy of the gradients of the trained POD modes resulting from the thermal gradients of the training thermal data, is the key to determine the accuracy of PODTherm-GP. This data-driven learning approach requires a very small number of POD modes (thus the DoF) to capture essential temporal and spatial thermal information embedded in the training data. If the quality of the training data is adequate, PODTherm-GP offers a very accurate prediction of dynamic thermal profile with a reduction in DoF by at least 5 orders of magnitude and a computing speedup by at least 3 orders of magnitude, compared to the FEM simulation. Practical for real-time thermal management, because only temperature in the high temperature areas at a certain time steps is needed, the saving in computational time for PODTherm-GP could be more than 4 orders of magnitude.   

Additionally, this work has illustrated the capability of PODTherm-GP beyond the training range.  Results show that PODTherm-GP offers an accurate prediction for dynamic thermal distribution in the selected 18-core CPU with a simulation time 40\% longer than the training time if more modes are included. Finally, an interesting finding reveals that PODTherm-GP could offer a more accurate prediction than the fine-resolution DNS if the training data are generated from finer-resolution DNS. If one can afford computational effort for the training and if the accuracy and efficiency for large-scale full-chip thermal simulation are the major concern, it is then worthwhile  to collect training thermal data in DNS with a very fine spatial resolution.

\bibliographystyle{ieeetr}
\vskip -2\baselineskip plus -1fil

\end{document}